\documentclass[twocolumn,showpacs,preprintnumbers,amsmath,amssymb]{revtex4}
\usepackage{graphicx}% Include figure files
\usepackage{dcolumn}% Align table columns on decimal point
\usepackage{bm}% bold math
\usepackage{epsfig}
\newcommand{\be}{\begin{eqnarray}}
\newcommand{\ee}{\end{eqnarray}}

\begin{document}

\title{1D multicomponent Fermions with delta function interaction in strong and weak
coupling limits:  $\kappa$-component Fermi gas }

\author{ Xi-Wen Guan$^{1}$, Zhong-Qi Ma$^{2}$ and  Brendan Wilson$^{1}$ }
\affiliation{$^{1}$ Department of
Theoretical Physics, Research School of Physics and Engineering,
Australian National University, Canberra ACT 0200, Australia}

%\author{ Zhong-Qi Ma}
\affiliation{$^{2}$ Institute of High Energy Physics, Chinese Academy
of Sciences, Beijing 100049, China}

\begin{abstract}

We derive the first few terms of the asymptotic expansion of the Fredholm
equations for one-dimensional $\kappa$-component fermions with repulsive and with attractive delta-function interaction in
strong and weak coupling regimes.  We thus obtain a highly accurate result  for the
ground state energy of a  multicomponent Fermi gas with
polarization for these regimes.  This result provides a unified
description of the ground state properties of the Fermi gas with higher
spin symmetries. However, in contrast to the two-component Fermi gas, there does not exist
a  mapping that can unify the two sets of Fredholm equations as the
interacting strength vanishes.  Moreover, we  find that  the local pair
correlation functions   shed light on the quantum statistic effects  of
the $\kappa$-component interacting fermions. For the  balanced spin case
with repulsive interaction, the  ground state energy obtained  confirms
Yang and You's result  [Chin. Phys. Lett. {\bf 28}, 020503 (2011)] that
the energy  per particle   as $\kappa \to \infty$ is the same as for
spinless Bosons.

\end{abstract}


\date{\today}
\maketitle

\section{Introduction}

The recently established experimental control over the effective spin-spin interaction between atoms is opening up new avenues for studying spin effects in low-dimensional atomic quantum gases with higher spin symmetries.  Fermionic alkaline-earth atoms display an exact $SU(\kappa)$ spin symmetry with  $\kappa= 2I + 1$ where $ I$ is the nuclear spin \cite{Gorshkov}. For example, a recent experiment \cite{Taie} for ${}^{171}$Yb dramatically realised the model of fermionic atoms with $SU(2) \otimes SU(6)$ symmetry where electron spin decouples from its nuclear spin $I = 5/2$. Such fermionic systems with enlarged $SU(\kappa)$ spin symmetry are expected to display a remarkable diversity of new quantum phases and quantum critical phenomena due to the rich linear and nonlinear Zeeman effects. The study of low-dimensional   cold atomic Fermi gases with higher pseudo-spin symmetries has become a new frontier in cold atom  physics.

One-dimensional (1D) quantum  Fermi gases with delta-function interaction  are  important exactly solvable quantum many-body systems and have had tremendous impact in quantum statistical mechanics.  The spin-1/2 Fermi gas with arbitrary polarization was solved long ago by Yang \cite{Yang}  using the Bethe ansatz (BA) hypothesis.   Sutherland \cite{Sutherland} generalized the result of the spin-1/2 Fermi gas  to 1D multi-component  Fermi gas in 1968.   The study of multicomponent attractive Fermi gases  was initiated   by  Yang \cite{Yang-a} and by Takahashi \cite{Takahashi-a}. Using Yang and Yang's method \cite{YangYang1969}
for the boson case, Takahashi \cite{Takahashi1971} and Lai
\cite{Lai1971} derived the  thermodynamic Bethe
ansatz (TBA) equations for spin-1/2 fermions.  In the same fashion, Schlottmann
\cite{Schlottmann} derived  the TBA equations for
$SU(\kappa)$ fermions with repulsive and attractive interactions
respectively.  Recently, the thermodynamics of the multicomponent Fermi gas was  obtained by solving  the TBA  equations  in  \cite{Guanb}. These models with enlarged spin symmetries have received a renewed interest in cold atom physics \cite{Controzzi,GuanPRL,Drummond,Jiang}.

In this communication,  we consider 1D $\kappa$-component fermions with repulsive and with attractive delta-function
interactions.  Although the  model was solved long ago by Sutherland \cite{Sutherland}, solutions of the Fredholm equations for the  model are far from being
thoroughly investigated except for the two-component Fermi gas \cite{guanma}.  From a theoretical point of view, finding a general  form of the  solutions to the Fredholm equations for   multicomponent Fermi gases with higher spin symmetry imposes a number of challenges.

In the present paper, using the method which we recently developed in the analytical study of  the 1D two-component Fermi gas \cite{guanma}, we approximately solve the Fredholm equations of the  1D $\kappa$-component fermions with polarization  for the A) strongly repulsive regime; B) weakly repulsive regime; C) weakly
attractive regime; and D) strongly attractive regime. We thus obtain the ground state energy of the $\kappa$-component Fermi gas with  polarization that provides a fundamental understanding of the ground state properties, such as phase diagram, magnetism and quantum statistical effects.  In contrast to the two-component Fermi gas \cite{guanma},  the two sets of Fredholm equations for the multi-component Fermi gas with  weakly repulsive and attractive interactions can not be unified   by the density mapping as the interacting strength vanishes.   The multiple spin degrees of freedom impose subtle intricacies of quantum statistics in the ground state properties of the systems. 
We further study  the local pair correlation functions for  the $\kappa$-component  interacting fermions. We find that the local pair correlation as $\kappa\to \infty$ is the same as for spinless bosons. This result is consistent with Yang and You's finding \cite{yangyou} that the energy  per particle   as $\kappa \to \infty$ is the same as for spinless Bosons.

\section{The Fredholm equations}

The Hamiltonian for the 1D $N$-body problem is \cite{Yang,Sutherland}
\begin{equation}
H=-\frac{\hbar^{2}}{2m}\sum_{i=1}^{N}\frac{\partial^{2}}{\partial
x_{i}^{2}}+g_{1D}\sum_{1\leq i<j\leq
N}\delta(x_{i}-x_{j}).\label{Ham}
\end{equation}
It describes $N$ fermions of the same mass $m$ confined to a 1D
system of length $L$ interacting via a $\delta$-function potential.
 There are 
$\kappa$ possible hyperfine states $|1\rangle, |2\rangle, \ldots,
|\kappa\rangle$ that the fermions can occupy. 
 For an irreducible representation $[\kappa^{N_{\rm \kappa}},(\kappa-1)^{N_{\rm \kappa-1}},\ldots,2^{N_{\rm 2}},1^{N_{\rm 1}}]$,   the Young diagram  has $\kappa$ columns with the quantum numbers  $N_i=N^{i}-N^{i+1}$,  here the $N^{\rm i}$ is the numbers of fermions at the $i$-th hyperfine levels  such that $N^{\rm 1}\geq N^{\rm 2}\geq\ldots\geq
N^{\kappa}$.  This system has
$SU(\kappa)$ spin symmetry and $U(1)$ charge symmetry.  The
coupling constant $g_{1D}$ can be expressed in terms of the
interaction strength $c=-2/a_{1D}$ as $g_{1D}=\hbar^{2}c/m$ where
$a_{1D}$ is the effective 1D scattering length. Here  $c>0$ for repulsive
fermions,  and $c<0$  for attractive fermions.

The energy eigenspectrum is given in terms of the quasimomenta
$\left\{k_i\right\}$ of the fermions via $E=\sum_{j=1}^Nk_j^2$, 
which in terms of the function $e_b(x)=(x+\mathrm{i}{bc}/{2})/(x-\mathrm{i}{bc}/{2})$ 
satisfy the BA equations  \cite{Sutherland,Yang-a,Takahashi-a}
\begin{eqnarray}
& &\exp(ik_{i}L)=\prod_{\alpha=1}^{M_{1}}e_1\left(k_i-\lambda_{\alpha}^{(1)}\right),\nonumber\\
&&\prod_{\beta=1}^{M_{\ell-1}}e_1\left(\lambda_{\alpha }^{(\ell)}-\lambda_{\beta}^{(\ell-1)} \right)\nonumber\\
&& =-\prod_{\eta=1}^{M_{\ell }}e_2\left(\lambda_{\alpha}^{(\ell)} -\lambda_{\eta}^{(\ell)}  \right) \prod_{\delta=1}^{M_{\ell+1}}e_{-1}\left( \lambda_{\alpha}^{(\ell)}-\lambda_{\delta}^{(\ell+1)}\right),
\label{BA}
\end{eqnarray}
Where $i=1,\ldots, N$, $\alpha=1,\ldots,M_{\ell}$ and 
the parameters $\left\{\lambda_{\alpha}^{(\ell)}\right\}$ with
$\ell=1,2,\ldots, \kappa-1$ are the spin rapidities while we denote 
$\lambda_{\alpha}^{(0)}=k_{\alpha}$,  $\lambda_{\alpha}^{(\kappa)}=0$ and $c'=c/2$.
The BA quantum numbers $M_i=\sum_{j=i}^{\kappa-1}(j-i+1)N_{\rm j+1}$ with
$M_{\kappa}=0$ in the above  BA equations (\ref{BA}).

\subsection{Repulsive regime} 

The fundamental physics of the model are determined by the set of transcendental BA equations (\ref{BA})  which can be transformed to generalised Fredholm equations in the thermodynamic limit. The Fredholm equations for repulsive and attractive regimes are significantly different.  From the BA equations (\ref{BA}), the quasimomenta $\left\{ k_i\right\}$ are real, but all  $\left\{ \lambda_{\alpha }^{(\ell)}\right\}$ are real only for  the ground state.  For the ground state,   the generalized Fredholm equations for $c>0$ are given by \cite{Sutherland,Yang-a,Takahashi-a}
\begin{eqnarray}
r_{0}(k)&=&\beta_{0}+\displaystyle
\int_{-B_{1}}^{B_{1}}K_{1}(k-k') r_{1}(k')dk',\label{FE0-R}\\
r_{m}(k)&=&\displaystyle \int_{-B_{m-1}}^{B_{m-1}}
K_{1}(k-\lambda)r_{m-1} (\lambda)d\lambda \nonumber\\
& &- \displaystyle \int_{-B_{m}}^{B_{m}}
K_{2}(k-\lambda) r_{m}(\lambda)d\lambda\nonumber \\
&& +\displaystyle \int_{-B_{m+1}}^{B_{m+1}}
K_{1}(k-\lambda) r_{m+1}(\lambda)d\lambda, \label{FE-R}
\end{eqnarray}
where $1\leq m \leq \kappa-1$  and $\beta_{0}=1/(2\pi)$, $r_{0}(k)$ is the particle
quasimomentum distribution function whereas $r_{m}(k)$ with $m\ge 1$
are the distribution functions for  the $\kappa-1$ spin  rapidities.
The kernel function $K_{\ell}(k)$ is defined as
\begin{equation}
K_{\ell}(k)=\displaystyle \frac{1}{2\pi}\displaystyle \frac{\ell
c}{(\ell c/2)^2+k^2},\label{K-r}
\end{equation}
here $c>0$ for  repulsive interaction and $c<0$ for attractive interaction. 
Following the method used for the two-component Fermi gas \cite{guanma},  we rewrite the Fredholm equations (\ref{FE-R})  as
\begin{eqnarray}
r_{m}(k)&=&\beta_{0} -
\sum_{s=0}^{m-1} \int_{|\lambda|>B_s}
K_{m-s}(k-\lambda) r_{s}(\lambda)d\lambda\nonumber\\
&& +
\int_{-B_{m+1}}^{B_{m+1}}K_{1}(k-\lambda)r_{m+1}(\lambda)
d\lambda ,\label{FE-R1}
\end{eqnarray}
where $0\leq m \leq \kappa-1$.
The associating integration boundaries $B_m$ are determined
by the conditions
\begin{equation}
m_\ell \equiv \frac{M_{\ell}}{L}=\int_{-B_{\ell}}^{B_{\ell}}{r_{\ell}}(k)dk,
\qquad 0\leq \ell  \leq \kappa-1,\label{B-R}
\end{equation}
where $M_{0}=N$ is the total number of fermions,
$N^{\ell }=M_{\ell-1}-M_{\ell}$ is the number of fermions in the $\ell$-th
hyperfine state. Here  $M_{\kappa}=0$. The ground
state energy $E$ per unit length is given  by
\begin{equation}
E=\int_{-B_0}^{B_0}k^2 r_{0}(k) d k. \label{E-R}
\end{equation}
The model has $SU(\kappa)$ symmetry in spin sector and $U(1)$ symmetry in charge sector. Therefore, the quantum numbers of each spin states are conserved.   Thus the system has $\kappa$ chemical potentials $\left\{\mu_\ell  \right\}$ in regard to these conserved numbers $N^{\ell }$. The ground state energy (\ref{E-R}) is a smooth function of the densities of $n^\ell=N^{\ell}/L$ with $\ell =1,2,\ldots, \kappa$ for unbalanced case.  In the grand canonical ensemble,  we can also get  the  chemical potentials  $\mu_\ell $ via $ \mu_\ell =\partial E/\partial n^\ell$.

\subsection{Attractive regime}

In the attractive regime, it is found that complex string solutions
of $k_{j}$ also satisfy the BA equations \cite{Yang-a,Takahashi-a,Gu-Yang}. The
quasimomenta $k_{j}$ may appear as bound states of $m$-atom up to length 
$2, \ldots, \kappa$. A bound state in quasimomentum space of length $m$ 
takes on the form 
\begin{equation}
k_{\alpha}^{m,j}=\lambda_{\alpha}^{(m-1)}+i(m+1-2j)|c'|+O(\exp(-\delta
L)), \label{attractiveroots}
\end{equation}
where $j=1,\ldots,m$. The number of  bound
states with length $1\leq m \leq\kappa$  is  denoted as  $N_{m}$.
Its real part is $\lambda_{\alpha}^{(m-1)}$. A
$k_{\alpha}$ bound state  of $m$-atom  will be accompanied by a
$\lambda_{\alpha}^{(1)}$ string of length $m-1$, a
$\lambda_{\alpha}^{(2)}$ string  of  length $m-2$ and so on until
a $\lambda_{\alpha}^{(m-1)}$ string  of length $1$. Each
accompanying  string  state in $\lambda^{(1)}$-space,
$\lambda^{(2)}$-space, \ldots, $\lambda^{(m-1)}$-space will share
the same real part $\lambda_{\alpha}^{(m-1)}$, see \cite{Yang-a,Takahashi-a,Gu-Yang} and  the second reference in \cite{Guanb}.  The unpaired atoms have  real quasimomenta $k_i$'s.

From these quasimomentum  bound states,   the Fredholm equations for the model with an attractive interaction   are given by  \cite{Yang-a,Takahashi-a}
\begin{eqnarray} 
\rho_{m}(\lambda)&=&m\beta_{0}+
\sum_{r=1}^{m-1} \sum_{s=r}^{\kappa}
\int_{-Q_{s}}^{Q_{s}}K_{s+m-2r}(\lambda-\Lambda) \rho_{s}(\Lambda)
d\Lambda\nonumber\\ 
&&+\sum_{s=m+1}^{\kappa} 
\int_{-Q_{s}}^{Q_{s}} K_{s-m}(\lambda-\Lambda)
\rho_{s}(\Lambda)d\Lambda,   \label{FE-A} 
\end{eqnarray}
where  $\rho_{1}(k)$ is the density distribution function of single fermions, whereas  $\rho_{m}(k)$ is  the density distribution function  for  the bound state  of $m$-atom  with $1<  m \le  \kappa$. The  total number
of fermions is given by $N = \sum_{m=1}^{\kappa} mN_m$. 
 The integration boundaries  $Q_m$,  characterizing the Fermi points in each Fermi sea, are determined by  
\begin{eqnarray}
  n_m\equiv \frac{N_m}{L}=\int_{-Q_m}^{Q_m}\rho_m(k)dk. \label{Nm-a}
\end{eqnarray}
The ground state  energy per unit length is given by
\begin{equation}
E=\sum_{m=1}^{\kappa}\int_{-Q_m}^{Q_m}\left(mk^{2}-\frac{m(m^{2}-1)}{12}c^{2}\right)\rho_{m}(k)dk.
\label{attractive-E}
\end{equation}

In the attractive regime, for convenience, we can define the effective chemical potentials for the cluster bound sates $\mu_m=\mu +H_m/m+(m^2-1)c^2/12$ with $m=1,\ldots, \kappa$, where $H_m$ is  the effective magnetic field (or Zeeman splitting parameter) for the bound  state of $m$-atom, see \cite{Guanb}.  The Zeeman energy per unit length can be written as $E_z=\sum_{m=1}^{N-1}H_mN_m/L$.   In this regime, the effective chemical potentials for the  bound states of different sizes can be derived from the energies of the ground state, i.e.
\begin{equation}
\mu_m=\frac{1}{m}\frac{\partial }{\partial n_m}\sum_{s=1}^{\kappa}\int_{-Q_s}^{Q_s}sk^2\rho_s(k)dk.\label{mu-a}
\end{equation}
Following \cite{Guanb}, the full phase diagrams of the model can be determined by the field-energy transfer relations
\begin{equation}
H_m=\frac{1}{12}m(\kappa ^2-m^2)c^2+m\left(\mu_m -\mu_{\kappa} \right) \label{E-H}
\end{equation}
with $m=1,2,\ldots, \kappa$. In the next section, we shall derive the explicit form of the ground state energy  that can  give highly accurate effective chemical potentials $\mu_m$ to determine magnetism and  full phase diagrams of the model.

Similarly, the Fredholm equations can be rewritten as 
\begin{eqnarray}
\rho_{m}(\lambda)&=&\beta_{0}-
\int_{|\Lambda|>Q_{m-1}} K_{1}(\lambda-\Lambda)
\rho_{m-1}(\Lambda) d\Lambda\nonumber \\
&&+\displaystyle \sum_{s=m+1}^{\kappa}\displaystyle
\int_{-Q_{s}}^{Q_{s}}K_{s-m}(\lambda-\Lambda) \rho_{s}(\Lambda)d\Lambda.  \label{FE-A1} 
\end{eqnarray}
We see that there is a particular mapping between the two sets of Fredholm equations  (\ref{FE-R1}) and
(\ref{FE-A1}),  i.e. 
\begin{eqnarray}
\rho_{m} &\rightarrow&  r_{\kappa-m},\qquad  Q_{m}\rightarrow B_{\kappa-m},\nonumber\\
\int_{-Q_{m}}^{Q_{m}} & \rightarrow  &\int_{-\infty}^{\infty}-\int_{-B_{\kappa-m}}^{B_{\kappa-m}},\qquad c \rightarrow -c .
\label{symmetry} 
\end{eqnarray}
This connection is  useful in the analysis of the symmetric structure between the two sides.  However, we find that the Fredholm equations (\ref{FE-R1}) for the repulsive regime  and  (\ref{FE-A1}) for the attractive regime do not preserve the mapping which exists between the two sets of the Fredholm equations for  the two-component Fermi gas \cite{guanma}. This leads to particular intricacies in  the analytical behaviour of the ground state energy as the interaction strength vanishes. 
In the present paper, we mainly concentrate on the  solutions of the two sets of the Fredholm equations for multicomponent Fermi gas with attractive and repulsive interactions.

\section{Asymptotic solutions of the Fredholm equations}

\subsection{Strong repulsion}

For the balanced case, i.e. $M_{m}-M_{m+1}=N/\kappa$, the following
theorem shows that all integration boundaries $B_{m}\to \infty$ except for $m=0$, namely, there are no finite Fermi points (without chemical biases between different spin states).\\

\noindent {\bf Theorem}:  For the $\kappa$-component fermion system with the repulsive delta-function interaction,  $B_{m}\rightarrow \infty$ for  $m=1,\ldots, \kappa-1$ if the relation 
\begin{equation}
M_{m-1}-M_{m}=M_{m}-M_{m+1} \label{B-B}
\end{equation}
holds, and   vice versa.

\noindent {\bf Proof}: Integrating both sides of the Fredholm equations  (\ref{FE-R}) with infinite boundary (where $m\neq 0$),  we obtain 
\begin{eqnarray}
&&  \int_{-\infty}^{\infty}r_{m}(k) dk
= \frac{M_{m-1}}{L}-\displaystyle \frac{M_{m}}{L}
+\frac{M_{m+1}}{L}.
\end{eqnarray}
If  the condition  (\ref{B-B}) holds, thus we have 
\begin{eqnarray}
\int_{-\infty}^{\infty}r_{m}(k) dk =\frac{M_{m}}{L}
\end{eqnarray}
where the integration boundary $B_m=\infty$ is inferred. Conversely, if
$B_{m}$ with $m\ge 1$  tend  to infinite,  from (\ref{FE-R}) and (\ref{B-R}) we have
\begin{eqnarray}
\frac{M_{m}}{L}=\displaystyle \int_{-\infty}^{\infty}
r_{m}(k) dk=\displaystyle \frac{M_{m-1}}{L}
- \frac{M_{m}}{L}+\frac{M_{m+1}}{L}. \nonumber  \qquad \diamond
\end{eqnarray}

\begin{figure}[tbp]
\includegraphics[width=1.0\linewidth]{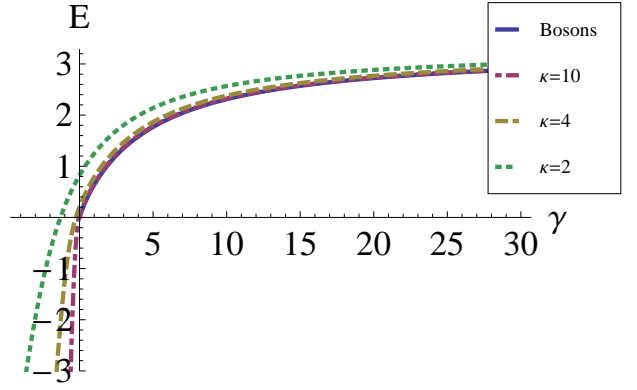}
\caption{ The ground state energy per length  vs  $\gamma=cL/N$ in natural   units of $ 2m=\hbar=1$ for the balanced  multicomponent Fermi gas: $\kappa =2, 4, 10$ and for spinless Bose gas in the whole interacting regime.  For the repulsive regime, the discrepancy between the energies of the $\kappa=10$ Fermi gas and  the spinless Bose gas is negligible. As $\kappa  \to \infty$, the ground state energy of the balanced repulsive Fermi gas fully  coincides with that of the spinless Bose gas \cite{yangyou}.  }
\label{fig:energy}
\end{figure}

For the  balanced case, the integration boundaries $B_m$ with $m\ge 1$ are infinitely large. The fermi momentum $B_0$ is always finite. For convenience, we introduce the notation $r_{0}(k)={r_0}_{\rm in}(k)+{r_{0}}_{\rm out}(k)$ where
\begin{eqnarray}
{r_0}_{\rm in}(k)&=&\left\{\begin{array}{ll}r_{0}(k)
\qquad \qquad &{\rm for }~~|k| \leq B_{0}\\ 0 &{\rm  for }~~|k| >B_{0}
\end{array} \right. \nonumber \\
{r_0}_{\rm out}(k)&=&\left\{\begin{array}{ll}0&{\rm for }~~|k| \leq B_{0}\\
r_{0}(k)\qquad \qquad  &{\rm for}~~|k| >B_{0}\end{array} \right. .\nonumber
 \end{eqnarray}
Taking the Fourier transformation with  the Fredholm equations (\ref{FE-R})
for the balanced case,  we obtain the following relations 
\begin{eqnarray}
\tilde{r}_{m}(\omega)&=&\displaystyle \frac{F_{m+1}(\omega)}{F_{m}(\omega)}
\tilde{r}_{m-1}(\omega),\,\,\, \tilde{r}_{1}(\omega)=\displaystyle \frac{F_{2}(\omega)}{F_{1}(\omega)}
\tilde{r}_{0_{\rm in}}(\omega),\nonumber\\
F_{m}(\omega)&=&\displaystyle \sum_{s=0}^{\kappa-m}
e^{-(\kappa-m+2s)c|\omega|/2},  \label{F-1}
\end{eqnarray}
where $2\leq m \leq \kappa-1$ and   $F_{\kappa}=1$. We denote the Fourier transform $\hat{F}[r_m(k)]=\tilde{r}_m(\omega)$. From the Fredholm equation (\ref{FE0-R}), we see that a closed form of the distribution function $r_1(k)$ is essential for the calculation  of the ground state energy. After a straightforward calculation with the relations   (\ref{F-1}), we obtain   the closed form
\begin{equation}
\tilde{r}_m(\omega)=\frac{\tilde{r}_{0_{\rm in}} (\omega)\sinh\left[ \frac{1}{2}(\kappa -m)|\omega |c\right]}{\sinh\left[\frac{1}{2}\kappa |\omega |c \right]}\label{r-m-o}
\end{equation}
that gives the distribution function 
\begin{equation}
r_{m}(\lambda )=\frac{1}{2\pi}\int_{-\infty}^{\infty}\tilde{r}_m(\omega) e^{-\mathrm{i} \omega \lambda }d\omega.\label{r-m}
\end{equation} 
In the above equations   $\tilde{r}_{0_{\rm in}}(\omega)$ can be determined  from the following relation
\begin{eqnarray}
I(k)&=&\int_{-B_{0}}^{B_{0}}K_{1}(k-\lambda)
r_{0}(\lambda)d\lambda\nonumber \\
&=&\int_{-\infty}^{\infty} K_{1}(k-\lambda)r_{0_{\rm in}}(\lambda)d\lambda,\label{I}  
\end{eqnarray}
where their Fourier transform reads $\tilde{I}(\omega)=e^{-c|\omega|/2} \tilde{r}_{0_{\rm in}}(\omega)$. 
The ground state energy per unit length (\ref{E-R}) can be calculated either numerically or asymptotically.  In Fig. \ref{fig:energy}, we  plot  the ground state energy of 1D balanced multicomponent  Fermi gas for $\kappa =2,\,4,\,10$ and for spinless Bose gas \cite{Lieb-Liniger}. 
We see that as $\kappa \to \infty$, the ground state energy of the balanced $\kappa$-component gas with repulsion  coincides with that of the 1D spinless Bose gas \cite{yangyou}.  In order to capture  the statistical nature for this connection, we shall calculate the first few terms of the expansion of the ground state energy.

The strong repulsion condition,  i.e.,  $cL/N\gg 1$, naturally gives $c\gg B_{0}$, where $B_{0}$  is proportional to the Fermi velocity  $k_F=n\pi$.  Following the method used in \cite{guanma},  we  take a Taylor expansion with respect to $\lambda$  for  the kernel function $K_{1}(k-\lambda)$ in eq.  (\ref{I}).   Using the relations (\ref{B-R}),  (\ref{E-R}), (\ref{r-m})  and   (\ref{I}), we  find 
\begin{eqnarray} 
\tilde{r}_{0_{\rm in}}(\omega)&\approx& n- \frac{E \omega^{2}  }{2},\nonumber\\
\tilde{r}_{1}(\omega)&\approx  &\frac{F_{2}(\omega)}{F_{1}(\omega)}
\left[ n- \frac{E \omega^{2}}{2}
\right]. \label{F-3}
\end{eqnarray}
In the above equations, we used the   following formulas 
\begin{eqnarray}
\hat{F}\left[ \left(c^{2}/4+k^{2}\right)^{-2} \right] &=&2\pi c^{-3}e^{-\frac{c|\omega|}{2}}\left[c|\omega|+2\right],\nonumber\\
\hat{F}\left[\left( c^{2}/4+k^{2}\right)^{-3} \right] &=& \pi c^{-5}e^{-\frac{c|\omega|}{2}}\left[c^{2}|\omega|^{2}+6c|\omega|
+12\right].\nonumber 
\end{eqnarray}

Substituting (\ref{F-3}) into the Fredholm equation  (\ref{FE0-R}), we
obtain an asymptotic form of the  density distribution function 
\begin{equation}
r_{0}(k)=\displaystyle \frac{1}{2\pi}
+ \frac{NY_{0}(k)}{2\pi L}-
\frac{EY_{2}(k) }{4\pi}+O(c^{-4}),\label{r-R}
\end{equation}
where the function
\begin{equation}
Y_\alpha (k)\approx \int_{-\infty}^{\infty}\frac{e^{\mathrm{i}\omega k}e^{-\frac{c|\omega|}{2} }\omega^{\alpha}F_2(\omega) }{F_1(\omega)}d\omega.
\end{equation}
After some algebra, we obtain  the two functions used  in (\ref{r-R})
\begin{eqnarray}
Y_0(k)&=&\frac{2Z_{1}}{c} - \frac{2Z_{3}
k^{2}}{c^{3}}+O(c^{-4})\nonumber\\
 Y_{2}(k)&=&\frac{4Z_{3}}{c^{3}}+O(c^{-4}),\nonumber
\end{eqnarray}
with
\begin{eqnarray}
Z_1&=&- \frac{1}{\kappa} \left[\psi(\frac{1}{\kappa})+C\right],\nonumber \\
Z_3&=& \kappa^{-3}\left[ \zeta(3,\frac{1}{\kappa})-\zeta(3)\right].\label{Z1-Z3}
\end{eqnarray}
Here $\zeta(z,q)$ and $\zeta(z)$ are  the Riemann  zeta functions,   $\psi(p)$ denotes the Euler psi function, $C$
denotes the Euler constant.
When $\kappa=2$
we have $Z_{1}=\ln~2,\,\,\,Z_{3}=(3/4)\zeta(3)$
that are consistent with the result given in  \cite{guanma}.

\begin{figure}[tbp]
\includegraphics[width=1.0\linewidth]{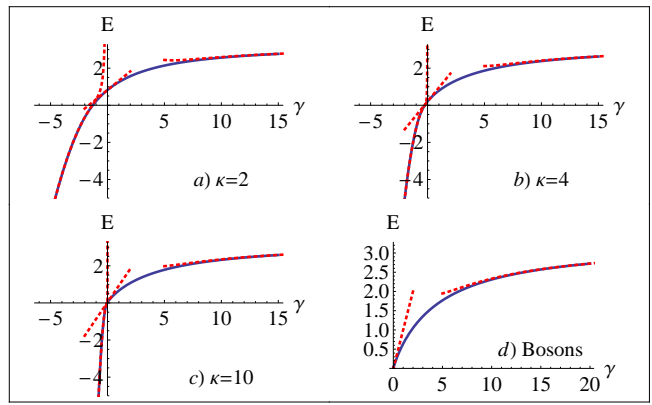}
\caption{ The ground state energy per length  vs  $\gamma=cL/N$ in natural  units of $ 2m=\hbar=1$ for the balanced  multicomponent Fermi gas: $\kappa =2, 4, 10$ and for spinless Bose gas.  The solid lines are the numerical solution  obtained from the two sets of Fredholm equations. The dashed lines are plotted directly from the asymptotic expansion ground state energies (\ref{E-R1}), (\ref{E-R5r}) for  a repulsive interaction and (\ref{E-AW}),  (\ref{E-A5-2}) for an attractive interaction.   High precision of the asymptotic expansion ground state energy is seen through out  strongly and weakly  coupled regimes.    }
\label{fig:E-table}
\end{figure}

Substituting (\ref{r-R}) into (\ref{B-R}) and (\ref{E-R}) and 
%\begin{eqnarray}
%n&=& \frac{B_{0}}{\pi}
%\left\{ 1+\displaystyle \frac{2nZ_{1}}{c }
%- \frac{2Z_{3}nB_{0}^{2}}{3c^{3} } -
%\frac{2Z_{3}E}{c^{3}}\right\}+O(c^{-4}), \nonumber\\
%E&=& \frac{B_{0}^{3}}{3\pi}\left\{
%1+ \frac{2nZ_{1}}{c }-
%\frac{6Z_{3}nB_{0}^{2}}{5c^{3} }-
%\frac{2Z_{3}E}{c^{3}}\right\}+O(c^{-4}). \nonumber
%\end{eqnarray}
solving by  iteration, the Fermi boundary $B_0$ and the ground state energy of balanced $\kappa$-component Fermi gas with a strong repulsion are given explicitly (up to order  $O(c^{-3})$)
\begin{eqnarray}
 B_{0}&\approx&   n\pi\left\{1-
\frac{2 Z_{1}}{\gamma}+\displaystyle \frac{4 Z_{1}^{2}}
{\gamma ^{2}}- \frac{8 Z_{1}^{3}}{\gamma ^{3}}+\frac{4Z_{3}\pi^{2}}{3\gamma ^{3}}
\right\},\\
E&\approx & \frac{n^3\pi^{2}}{3}\left\{1
-\frac{4 Z_{1}}{\gamma }+
\frac{12 Z_{1}^{2}}{\gamma^2 }- \frac{32 }{\gamma^3}\left(  Z_{1}^{3}-
\frac{Z_{3}\pi^{2}}{15}\right) \right\},\label{E-R1}
\end{eqnarray}
where the dimensionless interaction strength $\gamma =c/n$. 
The ground state energy  (\ref{E-R1}) with $\kappa =2$ reduces to the result given  in \cite{guanma}. 
In view of the quasi-momentum distribution $r_0(k)$ (\ref{r-R}), the system can be taken as an ideal gas with a  less exclusive  fractional statistics  than the  Fermi statistics.
It is interesting to see that the ground state energy (\ref{E-R1}) reduces to the energy of the spinless Bose gas  (see (10) in \cite{guan2}) as  $\kappa \to \infty$, also see  \cite{yangyou}. Here  we find that 
%\begin{equation}
$ \lim_{\kappa\rightarrow \infty}Z_{1}=
 \lim_{\kappa\rightarrow \infty}Z_{3}=1$.
 %\label{K-I}
%\end{equation}
 A significant interpolation between Fermions and Bosons can be conceived from the  parameters $Z_1$ and $Z_2$ in  (\ref{E-R1})  that encodes the  quantum statistical and dynamical effects.  In Fig. \ref{fig:E-table}, we see a good agreement between the asymptotic expansion result (\ref{E-R1}) and numerical result obtained from the Fredholm equation (\ref{FE0-R})  with the density distribution (\ref{r-m}).

In general, the boundaries $B_{m}$ with $m>1$ are hard to be estimated  for the imbalanced case in  a strong repulsive regime.  However, for  strong repulsion, the spin sector is dominated by the density-density interaction. Therefore, the spin polarization is not essential.  We consider  the high polarization case where $M_{m-1}-M_{m}\ll N$. In this case,  the condition $c\gg B_m$ always holds. Thus it is straightforward to calculate density distributions from (\ref{FE0-R})  and (\ref{FE-R}) with a proper Taylor expansion (up to order $O(c^{-3})$), i.e. 
\begin{eqnarray}
r_{0}(k)&\approx & \frac{1}{2\pi} +
\frac{2M_{1}}{c\pi L}\left[1- \frac{4k^{2}}
{c^{2}}\right], \nonumber\\
r_{1}(k)&\approx &\frac{1}{c\pi L}
\left[2N-M_{1}+2M_{2}\right]\nonumber\\
&&- \frac{k^{2}} {c^{3}\pi L}
\left[8N-M_{1}+8M_{2}\right]-
 \frac{8E}{c^{3}\pi}, \nonumber\\
r_{m}(k)&\approx & \frac{1}{c\pi L}\left[2M_{m-1}-M_{m}
+2M_{m+1}\right]\nonumber\\
&&
- \frac{k^{2}} {c^{3}\pi L}\left[
8M_{m-1}-M_{m}+8M_{m+1}\right],
\end{eqnarray}
 where $m=2,\ldots, \kappa$. It is also straightforward to calculate the Fermi boundary $B_0$ and $E$ from (\ref{B-R}) and (\ref{E-R}), i.e.
\begin{eqnarray}
B_{0}&=& n\pi\left\{1-
\frac{4m_{1}}{c }+\displaystyle \frac{16m_{1}^{2}}{c^{2}
}-\displaystyle \frac{64m_{1}^{3}}{c^{3}}\right.\nonumber\\
&& \left. +\frac{16\pi^{2}m_{1}n^{2}} {3c^{3}}
\right\}+O(c^{-4}),\nonumber\\ 
E&=&\frac{ n^{3}\pi^{2}}{3}\left\{1
- \frac{8m_{1}}{c }+
\frac{48m_{1}^{2}}{c^{2}}-
\frac{256m_{1}^{3}}{c^{3}}\right.\nonumber\\
&& \left.+ \frac{32\pi^{2}m_{1}n^{2}}
{5c^{3}}\right\}+O(c^{-4}),
 \label{E-HP} 
 \end{eqnarray}
 where $m_1=M_1/L$ with $M_1=\sum_{j=1}^{\kappa-1}N^{\rm j+1}$.
For the highly polarized  case, $M_{1}\ll N$, the ground
state energy $E$ per unit length, up to $O(c^{-3})$, solely depends on the quantum number $M_1$. The spin states  are  not essential in this strongly repulsive regime due to the freezing of  spin transportation.  This result agrees  with
the energy of the two-component Fermi gas with a strong repulsion,  see \cite{guanma}.
It  clearly indicates that strong repulsion suppresses the spin effect.

\subsection{Weak repulsion }

 For the balanced case with  weak repulsion, i.e., $cL/N\ll 1$,  all $B_{m}$
with $m\ge  1$  tend to infinity, so the Fredholm
equations (\ref{FE-R1})  become 
\begin{eqnarray}
r_{0}(k)&=&\beta_{0}+ 
\int_{-\infty}^{\infty}K_{1}(k-\lambda)r_{1}(\lambda)
d\lambda ,\nonumber\\ 
r_{m}(k)&=&\beta_{0}-  \int_{-\infty}^{\infty}
K_{m}(k-\lambda) r_{0_{\rm out}}(\lambda)d\lambda\nonumber \\
&&  +
\int_{-\infty}^{\infty}K_{1}(k-\lambda)r_{m+1}(\lambda)
d\lambda,\label{BA-w0}
\end{eqnarray}
where $1\leq m \leq \kappa-1$. By iteration with  $r_{1}$, $r_{2}, \ldots, r_{\kappa-1}$, the first equation of (\ref{BA-w0})  becomes 
\begin{equation}
r_{0}(k)=\kappa \beta_{0}-\displaystyle \sum_{s=1}^{\kappa-1}
\displaystyle \int_{-\infty}^{\infty} K_{2s}(k-\lambda)
{r_{0}}_{\rm out}(\lambda)d\lambda. \label{R0-R}
\end{equation}
Using  a Fourier transformation, we easily  prove
\begin{eqnarray}
r_{0_{\rm out}}(k)&=&\beta_{0}
- \int_{-B_{0}}^{B_{0}}R(k-\lambda) r_{0_{\rm in}}(\lambda)d\lambda,\nonumber\\
R(k)&=&\displaystyle \frac{1}{2\pi}
\int_{-\infty}^{\infty} e^{{\rm i}k\omega}
\left[ \sum_{s=0}^{\kappa-1} e^{-cs|\omega|}\right]^{-1}d\omega.
\end{eqnarray} 
Here we see that $R(k)=O(c)$.  Substituting the leading term of $r_{0_{\rm out}}(k)=\beta_{0}$ into
(\ref{R0-R}),  we further obtain 
\begin{eqnarray}
r_{0}(k)&=& \frac{\kappa }{2\pi}-\displaystyle \frac{1}{2\pi^{2}}\displaystyle \sum_{s=1}^{\kappa-1}\left\{\arctan\left(\displaystyle \frac{c s}{B_{0}-k}\right)\right.\nonumber\\
&& \left. + \arctan\left(\displaystyle \frac{cs}{B_{0}+k}\right)\right\}
+O(c^{2})  \label{R0-R2}
\end{eqnarray}
for the region  $|k|\leq B_{0}$. The energy can be calculated  from the equations (\ref{B-R}) and (\ref{E-R}) with  the distribution function (\ref{R0-R2})
%\begin{eqnarray}
% n&=& \frac{\kappa B_{0}}{\pi} \left\{1- \displaystyle \sum_{s=1}^{\kappa-1}\left[\displaystyle \frac{cs}{\pi \kappa B_{0}}\right.\right.\nonumber \\
%&& \left.\left.+ \frac{cs}{2 \pi \kappa B_{0}}\ln\left(\displaystyle \frac{c^{2}s^{2}+4B_{0}^{2}} {c^{2}s^{2}}\right)\right]\right\}+O(c^{2}),\nonumber\\
%E&=&\frac{\kappa B_{0}^{3}}{3\pi} \left\{1+\displaystyle \sum_{s=1}^{\kappa-1}\left[\displaystyle \frac{3c s}{\pi \kappa B_{0}}\right.\right.\nonumber\\
%&& \left.\left.-\frac{3cs}{2\pi \kappa B_{0}} \ln\left(\displaystyle\frac{c^{2}s^{2}+4B_{0}^{2}}{c^{2}s^{2}}\right)\right]\right\}+O(c^{2}).
%\end{eqnarray}
\begin{equation}
E=\displaystyle \frac{\pi^{2}n^{3}}{3\kappa^{2}} +c(\kappa-1)n^2/\kappa +O(c^{2}). \label{E-R0}
\end{equation}
When $\kappa=2$ the energy (\ref{E-R0}) reduces to the result   given  in \cite{guanma}. This result clearly indicates a mean field effect among the balanced $\kappa$-component weakly interacting fermions.  The kinetic energy part in (\ref{E-R0}) vanishes as $\kappa \to \infty$.   The  energy (\ref{E-R0})   as $\kappa \to \infty$ is the same as for spinless Bosons with a weak repulsion.

For the imbalanced case with weak repulsion, we assume 
$B_{m-1}> B_{m}$  with  $1\leq m \leq \kappa-1$.  The calculation of the ground state energy is very complicated because the $\kappa$ integral  equations  are coupled with each other. We have to separate the integration intervals case by case. 
Using  the Fredholm equations (\ref{FE-R1}), we calculate the 
following integral  in different regions.  We denote the integral $I_m= \int_{-B_{m}}^{B_{m}}
r_{m}(\lambda)K_{\ell}(k-\lambda)d\lambda$.  For the region  $|k|> B_{m}$, we find 
\begin{eqnarray}
I_m&=& \frac{\ell
cB_{m}}{2\pi^{2}(k^{2}-B_{m}^{2})}+\displaystyle
\sum_{s=m+1}^{\kappa-1}\displaystyle
\frac{cB_{s}}{2\pi^{2}(k^{2}-B_{s}^{2})}+O(c^{2}). 
\label{RK-1}
\end{eqnarray}
After a lengthy calculation,  we calculate the integral for  the region $|k|\leq B_{m}$
\begin{eqnarray}
 I_m&=&\beta_{0}-
\frac{\ell cB_{m}}{2\pi^{2}
(B_{m}^{2}-k^{2})} -\sum_{s=0}^{m-1}
\frac{(\ell -m+s+1)cB_{s}}{2\pi^{2}
(B_{s}^{2}-k^{2})}\nonumber \\
&&+ \int_{-B_{m+1}}^{B_{m+1}}
r_{m+1}(\lambda)K_{\ell+1}(k-\lambda)
d\lambda +O(c^{2}).\label{RK-2}
\end{eqnarray}
For  $B_{p}< |k|\leq B_{p-1}$ with $p>m$, we further calculate 
\begin{eqnarray}
 I_m&=&(p-m)\beta_{0}- \sum_{s=0}^{m-1}
\frac{(p-m)(\ell+s-m+1)cB_{s}}{2\pi^{2}(B_{s}^{2}-k^{2})}
\nonumber\\
&& - \sum_{s=m}^{p-1}\frac{[(p-s)(\ell+s-m+1)-1]
cB_{s}}{2\pi^{2} (B_{s}^{2}-k^{2})} \label{RK-4}\\
&&+  \frac{(\ell+p-m)cB_{p}}{2\pi^{2} (k^{2}-B_{p}^{2})}
+ \sum_{s=p+1}^{\kappa-1} \frac{cB_{s}}{2\pi^{2}
(k^{2}-B_{s}^{2})} +O(c^{2}).\nonumber
\end{eqnarray}
With the help of  these  formulas,  we are able to evaluate the order of $r_{m}(k)$ in the Fredholm equations 
(\ref{FE-R1})
\begin{eqnarray}
\left\{ \begin{array}{ll} 
r_{m}(k)=O(c) &  {\rm for } \,|k| > B_{m-1}\\
r_{m}(k)=\beta_{0}+O(c)  & {\rm for }\, B_{m+1}< |k| \leq B_{m-1}\\
r_{m}(k)=(p-m)\beta_{0}+O(c)& {\rm for }\, \begin{array}{c} B_{p}< |k| \leq B_{p-1}\\ p> m+1\end{array}
\end{array}
\right. .\label{RK-5}
\end{eqnarray}
From (\ref{B-R}), we obtain the integration boundaries via 
\begin{eqnarray}
\frac{M_{m}}{L}&=&
\frac{B_{m}}{\pi}+ \frac{M_{m+1}}{L}+
\frac{c}{2\pi^{2}}\left\{
\sum_{s=0}^{m-1}\ln\left|
\frac{B_{s}-B_{m}}{B_{s}+B_{m}}\right|\right.\nonumber\\
&&\left.+\sum_{s=m+1}^{\kappa-1}\ln\left|
\frac{B_{m}-B_{s}}{B_{m}+B_{s}}\right|\right\}
+O(c^{2}), \label{B-R1}
\end{eqnarray}
where $0\leq m \leq \kappa-1$. In order to calculate the ground state energy,  we also need a lengthy calculation of  the integral  
\begin{eqnarray}
&& \int_{-B_{m}}^{B_{m}}\lambda^{2} r_{m}(\lambda)
d\lambda = \frac{B_{m}^{3}}{3\pi}+
\frac{cB_{m}}{\pi^{2}}\left[
\sum_{s=0}^{m-1}B_{s}\right.\nonumber \\
&& \left.+ \sum_{s=m+1}^{\kappa-1}B_{s}\right]
+ \frac{c}{2\pi^{2}} \left\{
\sum_{s=0}^{m-1} B_{s}^{2}\ln\left| 
\frac{B_{s}-B_{m}}{B_{s}+B_{m}}\right|\right.\nonumber\\ 
&& \left.  +
\sum_{s=m+1}^{\kappa-1}B_{s}^{2}\ln\left|
\frac{B_{m}-B_{s}}{B_{m}+B_{s}}\right|\right\} \nonumber \\
&&+\displaystyle \int_{-B_{m+1}}^{B_{m+1}}\lambda^{2}
r_{m+1}(\lambda) d\lambda+O(c^{2}).
 \label{KRHO}
 \end{eqnarray}
Using (\ref{E-R}) and (\ref{KRHO}), the ground state energy $E$ per unit length 
is given by 
\begin{eqnarray}
E&=&
\sum_{m=0}^{\kappa-1}  \frac{B_{m}^{3}}{3\pi}
+ \frac{c}{2\pi^{2}} \sum_{m=0}^{\kappa-2}
\sum_{r=m+1}^{\kappa-1}\left[4B_{m}B_{r}\right.\nonumber \\
&& \left.+ \left(B_{m}^{2}+B_{r}^{2}\right)
\ln\left|\frac{B_{m}-B_{r}}{B_{m}+B_{r}}
\right|\right]+O(c^{2}).\label{E-wr}
\end{eqnarray}
Substituting  (\ref{B-R1}) into (\ref{E-wr}), we finally obtain the ground state energy of the $\kappa$-component Fermi gas with weak repulsion 
\begin{eqnarray}
E&=&\frac{1}{3}\sum_{i=1}^{\kappa-1} (m_{i-1}-m_i)^3\pi^2 \label{E-R5r} \\
&&+2c\sum_{i=1}^{\kappa-1}\sum_{j=i+1}^{\kappa}(m_{i-1}-m_i)(m_{j-1}-m_j)+O(c^{2}),
\nonumber
\end{eqnarray}
with $m_0=n$ and $m_{\kappa}=0$. Here the linear density $n=N/L$ and the quantum numbers $m_i=M_i/L$. 
If we introduce polarization $p_i=N^{\rm i}/L$ with $i=1,2,\ldots, \kappa-1$,  where $N^i$ is the number of fermions in the $i$th level. Thus we have a simple form of the ground state energy per length 
\begin{eqnarray}
E =\frac{1}{3}\sum_{i=1}^{\kappa}p_i^3\pi^2+2c \sum_{i=1}^{\kappa -1}\sum_{j=i+1}^{\kappa}p_ip_j+O(c^{2}).\label{E-R5r2}
\end{eqnarray}
The first part is the kinetic energy of the $\kappa$-component  fermions whereas the second parts is the interaction energy. 
This result is valid for arbitrary spin imbalance in the weakly repulsive regime. This result is in a good agreement with the numerical calculation, see Fig. \ref{fig:E-table}. For the balanced case, i.e. $M_{m}-M_{m+1}=N/\kappa$, (\ref{E-R5r}) reduces to the energy 
(\ref{E-R0}). It is interesting to note that the ground state energy (\ref{E-R5r2}) presents a mean field theory of the two-body s-wave scattering physics.

\subsection{Weak attraction}

In weakly attractive  coupling regime,  the Fredholm equations give the distribution functions of clusters  of different  sizes. 
For weak attraction, i.e.,  $|c|L/N\ll 1$,  the two sets of the Fredholm equations for repulsive and attractive regimes preserve  the symmetry (\ref{symmetry}).
Therefore, the calculation of the ground state energy $E$ per unit length for weak attraction is similar to that for weak repulsion.
For the balanced case with a  weak attraction, $N_{m}=0$ for $m=1,\ldots, \kappa-1$ and  $N_{\kappa}=N/\kappa$. Thus, we have the condition $Q_{m}=0$ for all Fermi points except   $Q_{\kappa}$.   In this case, the ground state is a spin singlet state. The Fredholm equations (\ref{FE-A}) for the spin neutral bound state of a $\kappa$-atom becomes
\begin{equation}
\rho_{\kappa}(\lambda)=\kappa \beta_{0}+\displaystyle \sum_{s=1}^{\kappa-1}\displaystyle \int_{-Q_{\kappa}}^{Q_{\kappa}}K_{2s}(\lambda-\Lambda)
\rho_{\kappa}(\Lambda)d\Lambda. \label{RK-A}
\end{equation}
Under the mapping (\ref{symmetry}),
the Fredhlom equations (\ref{RK-A}) for  weak attraction  map to  (\ref{R0-R}) for   weak repulsion. 
In order to estimate the contribution of density distribution $\rho_{\kappa_{\rm in}}$,  we take  the Fourier transformation of  (\ref{RK-A}) and then prove that 
\begin{eqnarray}
\rho_{\kappa_{\rm in}}(\lambda)&=&\beta_{0}- \int_{Q_{\kappa}}^{Q_{\kappa}}T(\lambda-\Lambda)
\rho_{\kappa_{\rm out}}(\Lambda)d\Lambda,\nonumber\\
T(\lambda)&=& \frac{1}{2\pi} \int_{-\infty}^{\infty}e^{{\rm i}\omega \lambda}
\left[ \sum_{s=0}^{\kappa-1}e^{-s|c\omega|}\right]^{-1}
d\omega,\nonumber
\end{eqnarray}
where $T(\lambda) =O(c)$. For $|\lambda|\leq Q_{\kappa}$, we thus have
$ \rho_{\kappa}(\lambda)={\rho_{\kappa}}_{\rm  in}(\lambda)
=\beta_{0}+O(c)$. 
Substituting this leading order into  (\ref{RK-A}), we have 
\begin{eqnarray}
\ \frac{n}{\kappa}&=&
\int_{-Q_{\kappa}}^{Q_{\kappa}}
\rho_{\kappa}(\lambda)d\lambda = \frac{Q_{\kappa}}{\pi}+
\frac{(\kappa-1)|c|}{2 \pi^{2}}\nonumber\\
&& +\displaystyle \sum_{s=1}^{\kappa-1} 
\frac{s|c|}{2\kappa \pi^{2}}\displaystyle \ln\left(
\frac{s^{2}|c|^{2}+4Q_{\kappa}^{2}}{s^{2}|c|^{2}}\right) +O(c^{2}).
 \label{N-A}
 \end{eqnarray}
Similarly, from (\ref{attractive-E}), we calculate the ground state energy of the balanced gas 
\begin{equation} 
E=  \frac{\pi^{2}n^{3}}{3\kappa^{2}}- \displaystyle
\frac{|c|(\kappa-1)n^{2} }{\kappa}
+O(c^{2}). \label{E-A-B} 
\end{equation}
We see that the energy of the balanced gas with  weak attraction  (\ref{E-A-B}) and with  weak  repulsion 
(\ref{E-R0}) continuously connect at $c\to 0$.  It is also seen that (\ref{E-A-B}) with $\kappa=2$ is consistent   with the result given in
\cite{guanma}. 

For the imbalanced case,  the ground state has cluster states of different sizes.  Therefore we assume an ansatz in order  to calculate the integration boundaries for each of the Fermi seas, i.e. 
$Q_{m}> Q_{m+1}$ with $1\leq m \leq \kappa-1$.
Similarly, we should calculate the integrals $I_m=\frac{1}{2\pi}\int_{-Q_m}^{Q_m}\frac{m|c|r_m(k)dk}{(mc/2)^2+(\lambda-k)^2}$  for different regions.  After some algebra, we find 
\begin{eqnarray}
I_m= \frac{\ell |c|Q_{m}}{2\pi^{2}
(\lambda^{2}-Q_{m}^{2})}  -
\sum_{s=m+1}^{\kappa}
\frac{|c|Q_{s}}{2\pi^{2}(\lambda^{2}-Q_{s}^{2})} +O(c^{2}), \label{rho-1}
\end{eqnarray}
for the region $|\lambda|> Q_{m}$  and 
\begin{eqnarray}
I_m&=&\beta_{0}-\displaystyle \frac{\ell |c|Q_{m}}{2\pi^{2}(Q_{m}^{2}-\lambda^{2})}
+\displaystyle \frac{\ell |c|Q_{m-1}}{2\pi^{2}(Q_{m-1}^{2}-\lambda^{2})}\nonumber\\
&& + \displaystyle \sum_{s=1}^{m-2}\displaystyle \frac{|c|Q_{s}}{2\pi^{2}(Q_{s}^{2}-\lambda^{2})}\label{rho-4} \\
&&  -\displaystyle \sum_{s=m+1}^{\kappa}\displaystyle \int_{-Q_{s}}^{Q_{s}}K_{\ell+s-m}(\lambda-\Lambda) \rho_{s}(\Lambda) d\Lambda+O(c^{2}). \nonumber
\end{eqnarray}
 for the region    $|\lambda|\leq Q_{m}$. In the region $Q_{m+1}< |\lambda|\leq Q_{m}$, we  further obtain  
\begin{eqnarray} 
&& I_m =\beta_{0}+  \sum_{s=1}^{m-2} \frac{|c|Q_{s}}{2\pi^{2}(Q_{s}^{2}-\lambda^{2})}+\frac{\ell |c|Q_{m-1}}{2\pi^{2}(Q_{m-1}^{2}-\lambda^{2})}\nonumber\\
 &&- \frac{\ell |c|Q_{m}} {2\pi^{2}(Q_{m}^{2}-\lambda^{2})}- \sum_{s=m+1}^{\kappa} \frac{(\ell+1)|c|Q_{s}}{2\pi^{2}(\lambda^{2}-Q_{s}^{2})}+O(c^{2}).
 \label{rho-5}
  \end{eqnarray}
For  $|\lambda|\leq Q_{m+1}$, we find 
\begin{eqnarray} 
I_m &=&\frac{(\ell+1)|c|Q_{m+1}} {2\pi^{2}(Q_{m+1}^{2}-\lambda^{2})}- \frac{(2\ell+1)|c|Q_{m}}{2\pi^{2}(Q_{m}^{2}-\lambda^{2})}\nonumber\\
&&+\frac{(\ell-1)|c|Q_{m-1}}{2\pi^{2} (Q_{m-1}^{2}-\lambda^{2})}+O(c^{2}).
\label{rho-6}
\end{eqnarray}

From these equations (\ref{rho-1}-\ref{rho-6}),  we are able to evaluate the leading order contributions in  the density distribution functions, i.e., 
$\rho_{m}(\lambda)= (m-p+1)\beta_{0}+O(c)$ 
for $Q_{p+1}< |\lambda|\leq Q_{p}$ with $p\leq m$  and
$\rho_{m}(\lambda)=O(c)$ for $|\lambda|> Q_{m}$. From equation (\ref{Nm-a}), we   obtain the integration boundaries
\begin{eqnarray} 
 \frac{N_{m}}{L}&=& \frac{Q_{m}}{\pi}- \sum_{s=m+1}^{\kappa}\ \frac{N_{s}}{L}-
\frac{|c|}{2\pi^{2}}\left\{ \sum_{s=1}^{m-1}\ln\left|
\frac{Q_{s}-Q_{m}}{Q_{s}+Q_{m}}\right|\right.\nonumber\\
&&\left.+\displaystyle \sum_{s=m+1}^{\kappa}\ln\left|
\frac{Q_{m}-Q_{s}}{Q_{m}+Q_{s}}\right|\right\}
+O(c^{2}). \label{B-A1} 
\end{eqnarray}
From (\ref{attractive-E}), the ground state energy $E$ per unit length is given by 
\begin{eqnarray}
E&=&\displaystyle \sum_{m=1}^{\kappa} \frac{Q_{m}^{3}}{3\pi}- \frac{2|c|}{\pi^{2}} \sum_{s=1}^{\kappa-1} \sum_{r=s+1}^{\kappa}Q_{s}Q_{r}\label{energy-wa}\\
&& - \frac{|c|}{2\pi^{2}} \sum_{s=1}^{\kappa-1} \sum_{r=s+1}^{\kappa}(Q_{r}^{2}+Q_{s}^{2})\ln\left|
\frac{Q_{s}-Q_{r}}{Q_{s}+Q_{r}}\right|+O(c^{2}).\nonumber 
\end{eqnarray}  
Substituting the integration boundaries $Q_m$  from (\ref{B-A1}) into (\ref{energy-wa}),   
we  thus  obtain the ground state energy of the $\kappa$-component Fermi gas with a weak attraction 
\begin{eqnarray}
E =\frac{1}{3}\sum_{i=1}^{\kappa} p_i^3\pi^2-2|c|\sum_{i=1}^{\kappa-1}  \sum_{j=i+1}^{\kappa} p_ip_j,\label{E-AW}
\end{eqnarray}
where we introduced polarization $p_i=N^{\rm i}/L$ with $i=1,2,\ldots, \kappa$.
The first part is the kinetic energy of the $\kappa$-component  fermions whereas the second part is the interaction energy.  This result is in a good agreement with numerical calculation, see Fig.\ref{fig:E-table}. For the balanced case, i.e. $N^{\rm i} =N/\kappa$, (\ref{E-AW}) reduces to the energy  (\ref{E-A-B}).
From the energies (\ref{E-R5r2}) and (\ref{E-AW}), we see that the ground state energy of the $\kappa$-component  gas with arbitrary polarization 
continuously connects at $c=0$, precisely speaking, the ground state energy per unit length and
its first derivative is continuously connected at $c=0$. Again, the ground state energy (\ref{E-AW}) presents a mean field theory of the two-body s-wave scattering physics in weak interacting regimes.

\subsection{Strong Attraction}

Universal low temperature behaviours of isospin $S = 1/2, 1, 3/2, \ldots, (\kappa-1)/2$ interacting fermions with an attractive interaction in 1D shed light on the nature of trions, pairing and quantum phase transitions.  The existence of these internal degrees of freedom gives rise to some exotic superfluid phases. These models exhibit new quantum phases of matter which are characterized by  bound states of different sizes underlying the symmetries. For example, three-component ultracold fermions give rise to a phase transition from a state of trions into the BCS pairing state under external fields. Recently, considerable interest has been paid to the low dimensional strongly interacting  fermionic atoms with high spin symmetries.

For a strong attraction, i.e. $|c|L/N\gg 1$, the bound states of different sizes form tightly bound molecules of different sizes with binding energies $\varepsilon_{\rm b}^{(\ell)}= \ell (\ell^{2}-1)c^{2}/12$, where $\ell=2,3,\ldots,\kappa$.  In this regime, all the Fermi momenta of the molecules are finite, i.e. $|c|\gg Q_m$ with $m=1,\ldots, \kappa$. Here $Q_1$ charaterizes the Fermi momentum of the single spin-alined  atoms.  In the canonical ensemble, external fields or nonlinear Zeeman splittings  trigger rich quantum phases and magnetism \cite{Guanb}.  A closed form of the ground state energy with polarization is essential to work out phase diagrams and magnetism at zero temperature.  In view of the strong attraction condition $|c|\gg Q_m$ and following the method \cite{guanma},  proper Taylor expansions can be carried out for the  three supplementary formulas
\begin{eqnarray}
&& \int_{-Q}^{Q}\bar{K}_{\ell}(\lambda-\Lambda)d\Lambda
= \frac{4Q}{\pi \ell |c|} -
\frac{16Q(Q^{2}+3\lambda^{2})}{3\pi \ell ^{3}|c|^{3}}\nonumber\\
&& +\frac{64Q(Q^{4}+10Q^{2}\lambda^{2}+5\lambda^{4})}{5\pi
\ell^{5}|c|^{5}}+O(c^{-7}),\nonumber\\
 && \int_{-Q}^{Q}(\lambda-\Lambda)\bar{K}_{\ell }(\lambda-\Lambda)d\Lambda\nonumber\\ 
&&=-\displaystyle \frac{4Q \lambda}{\pi \ell |c|} +
\frac{16Q \lambda(Q^{2}
+\lambda^{2})}{\pi \ell^{3}|c|^{3}}+O(c^{-5}),\nonumber\\
&&  \int_{-Q}^{Q}\Lambda^{2}\bar{K}_{\ell }(\lambda-\Lambda)d\Lambda\nonumber\\
&&=\displaystyle \frac{4Q^{3}}{3\pi \ell |c|}-\displaystyle
\frac{16Q^{3}(3Q^{2}+5\lambda^{2})}{15\pi \ell^{3}|c|^{3}}+O(c^{-5}).\nonumber
\end{eqnarray}
In the above  calculation, we denote the kernel function 
$\bar{K}_{\ell}(k)=\displaystyle \frac{1}{2\pi}\displaystyle \frac{\ell
|c|}{(\ell c/2)^2+k^2}.$ The following calculations are valid for arbitrary polarization. For a strong attraction, all terms in the Taylor expansion of the Fredholm equations are converged well, see  the method proposed in \cite{guanma}.   From  the Fredholm equations (\ref{FE-A}) and the boundary conditions (\ref{Nm-a}), it  is straightforward to obtain the following expression (up to the order $O(c^{-3})$)
\begin{eqnarray}
&& \displaystyle \frac{N_{m}}{L}
\approx \frac{mQ_{m}}{\pi}\left\{1- \frac{4}{mL|c|}F_{m}+ \frac{16\pi^{2}}{3m^{3}L^{3} |c|^{3}}G_{m}\right\},\nonumber \\
\label{rho-A1} 
&& \int_{-Q_{m}}^{Q_{m}} \Lambda^{2}\rho_{m}(\Lambda) d\Lambda\approx \displaystyle \frac{m Q_{m}^{3}}{3\pi}\left\{1-
\displaystyle \frac{4}{mL|c|}F_{m}\right.\nonumber\\ 
&& \left.+
\frac{16\pi^{2}}{15m^{3}L^{3}|c|^{3}}\overline{G}_{m}\right\},  \label{E-A6} 
\end{eqnarray}
where we denoted the functions 
\begin{eqnarray} 
F_{m}&=& \sum_{s=1}^{m-1}
 \sum_{r=1}^{s} \frac{N_{s}}{ 2r-s+m-2}+ \sum_{r=1}^{m-1} \frac{N_{m}}{2r}\nonumber\\
&& + \sum_{s=m+1}^{\kappa} \sum_{r=1}^{m} \frac{N_{s}}{ 2r+s-m-2},\nonumber\\
G_{m}&=&\displaystyle \sum_{s=1}^{m-1}
\sum_{r=1}^{s} \frac{s^{2}N_{m}^{2}N_{s} +m^{2}N_{s}^{3}}{s^{2} (2r-s+m-2)^{3}}+ \sum_{r=1}^{m-1}
\frac{2N_{m}^{3}}{(2r)^{3}}\nonumber\\
&& +\sum_{s=m+1}^{\kappa}
 \sum_{r=1}^{m} \frac{s^{2} N_{m}^{2}N_{s}+m^{2}N_{s}^{3}}{s^{2} (2r+s-m-2)^{3}},\nonumber \\
\overline{G}_{m}&=&  \sum_{s=1}^{m-1}
\sum_{r=1}^{s} 
\frac{9s^{2}N_{m}^{2}N_{s} +5m^{2}N_{s}^{3}}{s^{2} (2r-s+m-2)^{3}}+ \sum_{r=1}^{m-1}
\frac{14N_{m}^{3}}{(2r)^{3}}\nonumber\\
&& + \sum_{s=m+1}^{\kappa}
 \sum_{r=1}^{m}
\frac{9s^{2} N_{m}^{2}N_{s}+5m^{2}N_{s}^{3}}{s^{2} (2r+s-m-2)^{3}}.\nonumber
 \end{eqnarray}
 Here $N_i$ with $i=1,2\ldots, $ are the numbers of the cluster state of $i$-atom. 

After a lengthy calculation, we obtain explicit forms of the Fermi momenta and the energies of the molecules of different sizes  for a strong attraction
\begin{eqnarray} 
 Q_{m}&\approx & \frac{N_{m\pi }}{mL}
\left\{1+ \frac{4}{mL|c|}F_{m}+ \frac{16}{m^{2}L^{2}|c|^{2}}F_{m}^{2}\right.\nonumber \\
&& \left.+\displaystyle \frac{16}{3m^{3}L^{3} |c|^{3}}\left[12F_{m}^{3}-G_{m}\pi^{2}\right]\right\},  \label{rho-A3} \\
E_{m}&\approx &\frac{\pi^{2}N_{m}^{3}}{3m L^{3}}
\left\{1+ \frac{8}{mL|c|}F_{m}\right.\nonumber\\ 
&& + \frac{48}{m^{2}L^{2}|c|^{2}}F_{m}^{2}+ \frac{256}{m^{3}L^{3}|c|^{3}}F_{m}^{3}\nonumber\\ 
&& \left.+\frac{16\pi^{2}}{m^{3}L^{3} |c|^{3}}\left[-G_{m}+\overline{G}_{m}/15\right]\right\}.  \label{E-A5} 
\end{eqnarray}
Thus the ground state energy of the gas with arbitrary polarization per length  in strong attractive regime is given by 
\begin{equation}
E=\sum_{\ell=1}^{\kappa}(E_\ell-n_\ell \varepsilon_{\rm b}^{(\ell)}), \label{E-A5-2}
\end{equation}
where the binding energy of the molecule state of $\ell$-atom is given by $\varepsilon_{\rm b}^{(\ell)}=\ell (\ell^2-1)c^2/12$.
For $\kappa=2$, it  covers the result obtained for the two-component Fermi gas given in  \cite{guanma}.  From the discrete BA equations,  one of the author and coworkers \cite{Guanb} derived the ground state  energy of $\kappa$-component strongly attractive  Fermi gas for up to the order $O(1/c^2)$.  Here we obtained    a more accurate ground state energy (\ref{E-A5}) of the $\kappa$-component gas  with arbitrary polarization  from the analytical study of the Fredholm equations.  We noticed that  the ground state of strongly attractive $\kappa$-component  Fermi gases can be effectively described by a super Tonks-Girardeau gaslike state via a proper mapping \cite{Yin}.
The explicit form of the ground state energy can be used to study  magnetism and  full phase diagrams of the model  from  the relations (\ref{mu-a}) and (\ref{E-H}) in a straight forward way. The effective chemical potentials   may be calculated from $\mu_m=\partial \left( \sum_{\ell=1}^\kappa E_\ell \right)/\left(m \partial n_m\right)$,  where $n_m=N_m/L$.  From these effective fields, the phase diagrams can be determined by the energy-field transfer relations (\ref{E-H}), see recent study of the three-component attractive Fermi gas \cite{GuanPRL,Angela} . 

\section{Local pair correlations}

There has been a considerable interest in studying  universal nature of interacting fermions. Remarkably, Tan \cite{Tan} showed that the momentum distribution exhibits universal ${\cal C}/k^4$ decay as the momentum tends to infinity. Here the constant ${\cal C}$ is called universal contact that measures the probability of two fermions with opposite spin at the same position. For  1D two-component Fermi gas \cite{Zwerger2}, the universal contact is obtained by calculating the change of the interacting energy with respect to interaction strength by Hellman-Feynman theorem, i.e. ${\cal C}=\frac{4}{a_{1D}^2}n_{\uparrow}n_{\downarrow}g^{(2)}_{\uparrow,\downarrow}(0)$. The local pair correlation $g^{(2)}_{\uparrow,\downarrow}$ are accessible via exact Bethe ansatz solution through the relation  $g^{(2)}_{\uparrow,\downarrow}(0) =\frac{1}{2n_{\uparrow}n_{\downarrow}}\partial E/\partial c$. Here $E$ is the ground state energy per length.

 In a similar way, for 1D $\kappa$-component Fermi gas, there exists a 1D analog of the Tan adiabatic theorem where the universal contact is given by the local pair correlations for two fermions with different spin states. The two-body local pair correlation function is similar to the calculation of the expectation value of the four-operator  term in the second quantized Hamiltonian.  For a homogenous and balanced $\kappa$-component Fermi gas, the local pair correlation function  is given by 
 \begin{equation}
 g^{(2)}_{\sigma,\sigma'}(0)=\frac{\kappa}{(\kappa-1)n^2}\frac{\partial E}{\partial c} \label{g2}
 \end{equation}
with $\kappa>1$.  Where $E$ is the ground state energy per length. From the asymptotic expansion result of the ground state energy of the balanced Fermi gas obtained in the above section, we easily find 
$g^{(2)}_{\sigma,\sigma'}(0)\to 1$ as $|c|\to 0$.  From the ground state energy (\ref{E-R1}) of the balanced gas with a strong repulsion,  we have the local pair correlation
\begin{eqnarray}
 g^{(2)}_{\sigma,\sigma'}(0)=\frac{4\kappa\pi }{3(\kappa-1)\gamma^2}\left[ Z_1-\frac{6Z_1^2}{\gamma }+\frac{24}{\gamma^2}\left( Z_1^3-\frac{Z_3\pi^2}{15}\right)\right]. \label{g2-rs}
\end{eqnarray}
This local pair correlation reduces to the one for the spinless Bose gas \cite{g2,guan2} as $\kappa \to \infty$.

\begin{figure}[tbp]
\includegraphics[width=1.050\linewidth]{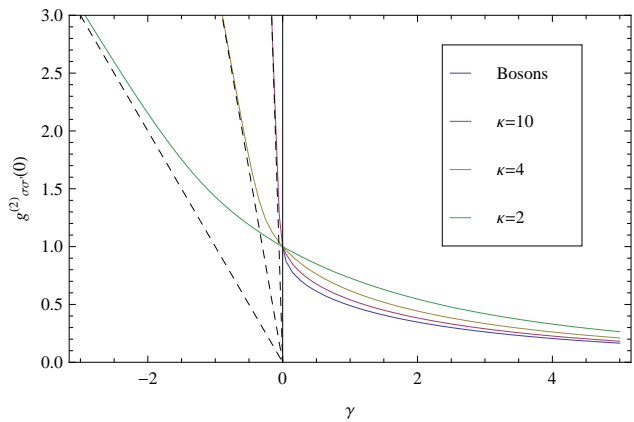}
\caption{ The local pair correlation $g^{(2)}_{\sigma,\sigma'}(0)$  vs  $\gamma=cL/N$ in natural  units for the balanced  multicomponent Fermi gas: $\kappa =2, 4, 10$ and for spinless Bose gas.  The solid lines are the numerical solutions  obtained from the two sets of Fredholm equations. The dashed lines are  the asymptotic limits obtained from the first term in (\ref{g2as}). All local correlations pass the point  $g^{(2)}_{\sigma,\sigma'}(0)=1$  at vanishing interaction strength.  However, the local pair  correlation diverges for the attractive Bose gas. }
\label{fig:g2}
\end{figure}

\begin{figure}[tbp]
\includegraphics[width=1.0\linewidth]{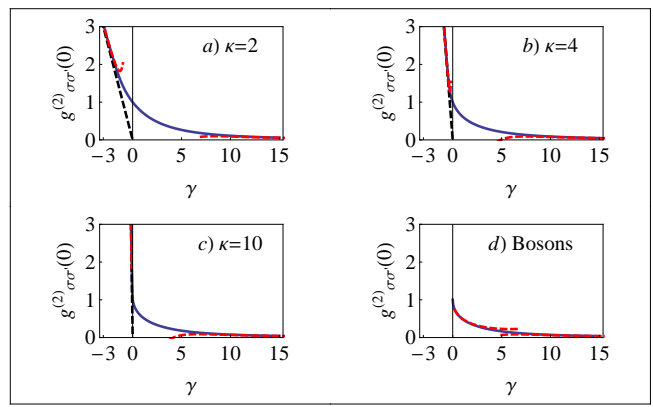}
\caption{ The local pair correlation $g^{(2)}_{\sigma,\sigma'}(0)$  vs  $\gamma=cL/N$ in natural  units for the balanced  multicomponent Fermi gas: $\kappa =2, 4, 10$ and for spinless Bose gas.  The solid lines are the numerical solutions obtained from the two sets of Fredholm equations. The dashed lines are  the asymptotic expansion result  obtained from  (\ref{g2-rs}) and (\ref{g2as}).  }
\label{fig:g2-table}
\end{figure}

For strong attractive interaction, we obtain the ground state energy of the balanced gas from the result (\ref{E-A5})
\begin{eqnarray}
E&=&\frac{\pi^2n^3}{3\kappa^4}\left(1+\frac{4A_\kappa}{\kappa^2 |\gamma|} +\frac{12A_\kappa^2}{\kappa^4\gamma^2}+\frac{32A_\kappa^3}{\kappa^6|\gamma|^3}\right. \nonumber\\
&& \left.-\frac{32B_\kappa}{15\kappa^6\gamma^3} \right)-n_\kappa \varepsilon_{\rm b}^{(\kappa)} +O(1/\gamma^4),
\end{eqnarray}
where $A_\kappa=\sum_{r=1}^{\kappa-1}1/r$ and $B_\kappa=\sum_{r=1}^{\kappa-1}1/r^3$.
From the relation (\ref{g2}), we obtained the local pair correlation for two fermions with different spin states in a strong attractive regime
\begin{eqnarray}
g^{(2)}&=&\frac{\kappa(\kappa+1)|\gamma|}{6}+\frac{4\pi^2}{3\kappa^5(\kappa-1)\gamma^2}\left(A_\kappa+\frac{6A_{\kappa}^2}{\kappa^2|\gamma|}\right.\nonumber\\
&&\left. +\frac{24}{\kappa^4 \gamma^2}\left(A_\kappa^3-\frac{B_\kappa}{15}\right)\right)+O(1/\gamma^4).\label{g2as}
\end{eqnarray}
We see that the local pair correlation becomes divergent as $\kappa \to \infty$, i.e. goes to the limit for  the attractive bosons. In Fig.\ref{fig:g2}, we present the numerical solution of the local pair correlation for two fermions with different spin states in the multicomponent Fermi gas. In the whole interacting regime, the local  pair correlations for  the balanced $\kappa$-component Fermi gas tends to the limit value for  the spinless bosons as $\kappa \to \infty$. High precision of these asymptotic expansion local pair correlations is seen in strong and week interaction regions, see Fig.\ref{fig:g2-table}.

\section{Conclusion}

We have analytically studied the Fredholm equations for the 1D $\kappa$-component fermions with repulsive and attractive $\delta$-function
interactions in four regimes: a) strong repulsion; B) weak repulsion; C) weak attraction and D) strong attraction. Solving   two sets  of the Fredholm equations
(\ref{FE-R}) and (\ref{FE-A}),  we  have  obtained the first few terms of the asymptotic expansion for the density distribution functions and 
the ground state energy $E$ per unit length for  both balanced and unbalanced cases in  these regimes. 
We summarize our main result as following.

A) For the strong repulsive regime, the ground state energy of the balanced $\kappa$-component  gas 
has been given in  (\ref{E-R1})  up to the order of $1/c^3$. It presents a universal structure in terms of the parameters $Z_1$ and $Z_3$ (\ref{Z1-Z3}) that characterize quantum statistical and dynamical effects of the model.  It   also confirms Yang and You's result  \cite{yangyou} that the energy   per particle as $\kappa \to \infty$ is the same as for spinless Bosons. In this regime, we have also obtained the ground state energy for the highly  imbalanced case (\ref{E-HP}). It clearly indicates that the internal spin effect is strongly suppressed in this regime, i.e. the energy only depends on the quantum number $M_1$.

B) For the weak repulsive regime, we have calculated the ground state energy $E$ (\ref{E-R0})
for the balanced case and  (\ref{E-R5r}) for the imbalanced case.  The ground state energy (\ref{E-R5r}) presents a simple mean field theory of the two-body s-wave scattering physics. 
We see that the energy  (\ref{E-R0}) for the balanced $\kappa$-component Fermi gas  is the same as that for  the spinless Bose gas as $\kappa \to \infty$, consistent with the result in \cite{yangyou}.

C) For the weak attractive regime, we have also calculated  the ground state energy $E$ (\ref{E-A-B})
for the balanced case and (\ref{E-AW})  for the imbalanced case.  We have found that  the two sets of the Fredholm equations for the model with repulsive and attractive interactions  preserves a particular mapping  (\ref{symmetry}). But they do not preserve the density mapping found for the two-component Fermi gas \cite{guanma}.  Our result shows that the ground state energy of the model with arbitrary spin state   is continuous at $c\to 0$, as is  the first derivative.

D) For the strong attractive regime, the highly accurate  ground state energy (up to $O(c^{-3})$) of the $\kappa$-component with arbitrary spin polarization   has been given in  (\ref{E-A5}). The ground state of the  largest cluster state can be effectively described by the gas like state in the super Tonks-Girardeau gas of hard core bosons.   From the  result of ground state energy of the strongly attractive fermions, we can study magnetism and identify full phase diagrams of the gas with  $SU(\kappa)$ symmetry driven by the  external fields.

Furthermore,  for the balanced case, we have obtained the first few terms of the asymptotic expansion for the local pair correlations of  two fermion with different spin states in the four regimes. A numerical solution  of the local pair correlation function  has been obtained in the whole interacting regime. These results we obtained   provide  insight  into understanding quantum statistical and dynamical effects in 1D interacting fermions with high spin symmetries. The explicit forms of the ground state energy of the multicomponent Fermi gas with polarization give further applications to the study of quantum phase transitions, magnetism, and phase diagrams, which provides useful guides  for future experiments with ultracold atomic Fermi gases with high symmetries.

\begin{acknowledgments}
The authors thank Professor Chen Ning Yang for initiating this topic and for  helpful discussions and suggestions.   This work is partly supported by the Australian Research Council and  by  Natural Science Foundation of China under the grants No. 11075014 and No. 11174099.
\end{acknowledgments}

\end{document}